\documentstyle[eqsecnum,aps]{revtex}
\input{epsf}

\newcommand{\be}{\begin{eqnarray}}
\newcommand{\ee}{\end{eqnarray}}
\newcommand{\bi}{\bibitem}

\begin{document}
\draft
\title{Classification of
singular points in polarization field of CMB\\ 
and eigenvectors of Stokes matrix.}
\author{A.D. Dolgov\cite{add}, A.G. Doroshkevich}
\address{Teoretical Astrophysics Center,
 Juliane Maries Vej 30, DK-2100, Copenhagen, Denmark}
\author{D.I. Novikov\cite{din}}
\address{Dept. of Physics and Astronomy, University of Kansas,
  Lawrence, Kansas, 66045}
\author{I.D. Novikov\cite{idn}}
\address{Teoretical Astrophysics Center,
 Juliane Maries Vej 30, DK-2100, Copenhagen, Denmark}
\date{\today}
\maketitle
\begin{abstract}
\setlength{\baselineskip}{0.3in}
Analysis of the singularities of the polarization field of CMB, where
polarization is equal to zero, is presented. It is found that the
classification of the singular points differs from the usual three
types known in the ordinary differential equations. The new statistical
properties of polarization field are discussed, and new methods to
detect the presence of primordial tensor perturbations are indicated.
\end{abstract}


\section{Introduction}

In the coming years the measurements of the angular
anisotropies of the intensity of cosmic microwave
background (CMB) by the cosmic missions MAP and PLANCK
will possibly present one of the most
promising methods of studying early universe as well as of precise
measuring of basic cosmological parameters(see e.g.\cite{revcmb} and
references therein). In addition to the anisotropies of the intensity,
it is possible, though more difficult, to measure polarization
of the radiation. The polarization is a secondary effect induced by the
scattering of anisotropic radiation field on electrons in cosmic plasma.
The corresponding measurements of the CMB polarization are planning to be
performed in the coming space missions 

Polarization field is described by a traceless $2\times 2$-matrix which
can be decomposed into a sum of three Pauli matrices $\sigma_i$
with the coefficients known as Stokes parameters:
\be
\mathbf{a} = \xi_i \sigma_i
\label{asigma}
\ee
As is well known circular polarization does not arise
in Thomson scattering (because of parity conservation),
so that $\xi_2 =0$ and the matrix
$\mathbf{a} $ is symmetric. Usually it is parameterized in the form:
\be
\mathbf{a} =
\left( \begin{array}{cc}
Q & U \\
U & - Q
\end{array}
\right)
\label{aqu}
\ee

The sources of polarization are anisotropies of radiation field induced
by different types of perturbations, namely
scalar, tensor and vector ones. Vector perturbations decay in the early
universe but may arise at small scales at later stages and influence the
CMB polarization in the case of reionization.

Geometrical properties of the polarization field allow to obtain an important
cosmological information [2-9]. An importance of
the study of singular points of polarization field, where $Q=U=0$,
was emphasized in ref. \cite{nn}. In this paper
geometric classification and statistics of the singular point was proposed.
This has been done in terms of the fields $Q$ and $U$ which directly
enter polarization matrix. The latter are 2-dimensional tensor fields and have
the appropriate transformation properties under rotation of coordinate
system.

\section{Types of singular points.}

In this paper we will investigate the classification of singular points
of eigenvectors of the polarization matrix ${\bf a}$. Though the positions
and statistics of the singular points remains the same, their types could be
quite different. The eigenvalues are easily
found:
\be
\lambda_{1,2} =\pm \sqrt { Q^2 +U^2}
\label{lambda12}
\ee
and the eigenvector corresponding to the positive $\lambda$ is
\be
\vec n^{(1)} \equiv \left( n_x, n_y \right)
\sim \left(U,\, \sqrt { Q^2 +U^2} - Q \right)
\label{n1}
\ee
This vector determines direction of maximum polarization and up to a
normalization factor coincides with the direction of the  vector $\vec P$
considered in refs. \cite{be,kks,zs} or orthogonal to it, depending upon the
sign of the coefficient functions.

The behavior of the vector field $\vec n\,^{(1)}$ in the vicinity of the
singular points of polarization, $Q=U=0$, is determined by the equation
\be
{dy \over dx} =  {n_y \over n_x} = {\sqrt { Q^2 +U^2} - Q \over U}
\label{dxdy}
\ee
Analysis of singular points of differential equations when both numerator
and denominator can be expanded into Taylor series is well known and
can be found e.g.in ref. \cite{bs}. In the usual case the following singular
points can exist: focus, saddle, and knot. In our case the situation is more
complicated due to the presence of the square root in the numerator
which is generically non-analytic in the points where $Q=U=0$.
At this stage a question
may arise why it is assumed that $Q$ and $U$ are analytic functions
expandable into Taylor series (at least up to first order terms) around
the points where $Q=U=0$, while the component of the eigenvectors are not.
The reason for that is the following. The matrix elements of the
polarization matrix $Q$ and $U$ are directly related to the anisotropy of the
cosmic microwave radiation through the amplitude of photon-electron
scattering. We do not expect any grounds for these quantities to have a
square root singularity where their first derivative tends to infinity.
On the other hand the eigenvectors of the matrix ${\bf a}$ just mathematically
contain the square root $\sqrt { Q^2 +U^2}$ and so it is singular at $Q=U=0$.
The analysis of singularities of the vector field $\vec n\,^{(1)}$
can be done as follows.
We assume that $Q$ and $U$ are expanded near singular points as
\be
Q= a_1 x + a _2 y    \nonumber \\
U = b_1 x + b_2 y
\label{qu}
\ee
In the case when the matrix
\be
 {\bf M} = \left( \begin{array}{cc}
a_1 & a_2 \\
b_1 & b_2
\end{array}
\right)
\label{m}
\ee
is not degenerate, det ${ \bf M} \neq 0$, it is convenient to introduce the
new coordinates:
\be
\xi= a_1 x + a _2 y,\,\,\,   \eta = b_1 x + b_2 y  \nonumber \\
 x=A_1 \xi +A_2 \eta,\,\,\, y =B_1 \xi +B_2 \eta
\label{xieta}
\ee
Evidently the types of the singular points do not change under this coordinate
transformation. It is simpler to make the further analysis in
polar coordinates:
\be
\xi=\rho \cos \phi,\,\,\, \eta=\rho \sin \phi
\label{rphi}
\ee

Equation (\ref{dxdy}) in this new polar coordinates can be rewritten as
\be
{d (\ln \rho ) \over dt } = {2\over t^2 +1} {N\over D}
\label{drdt}
\ee
where $t= \tan (\phi /2)$ and
\be
N= -A_2 t^3 + t^2 (B_2 -2A_1) +t(2B_1 +A_2) -B_2
\label{n}
\ee
and
\be
D= A_1 t^3 - t^2 (B_1 + 2A_2) +t(2B_2 - A_1) +B_1
\label{d}
\ee
Barring the degenerate case of $A_1 =0$ we may take $A_1 =1$ without loss of
generality.

The behavior of singular points depends upon the roots of denominator $D$.
Let us first consider the case when it has three real root, $t_{1,2,3}$.
The solution of eq. (\ref{drdt}) in this case can be written as:
\be
{r\over r_0} = \left( t^2 +1 \right)
\prod_j \mid t-t_j\mid^{2\nu_j}
\label{r3}
\ee
where $r_0$ is an arbitrary constant and
the powers $\nu_j$ are
\be
\nu_1 = -{ (1+t_1^2)(1+t_2 t_3) \over 2(t_1-t_2)(t_1-t_3)}
\label{nu1}
\ee
and so on by cyclic permutation of indices. Since it can be shown that
$\sum \nu_j = -1$, the
points where $t^2\rightarrow \infty$ are not generally singular.
It can be easily checked that
either all $\nu_j <0$ or any two of them are negative and one is positive.
In the first case the singular point resembles the usual saddle with the
only difference that there are three and not four, as in the usual case,
linear asymptotes/separatrices (see Fig.1a). We will call it also "saddle".
If one of $\nu_j$ is positive
(say $\nu_1 > 0$), and thus $r$ becomes zero at $t=t_1$, the behavior of
the direction field near this point is quite different from the usual
ones. The field line cannot be continued along $\phi =\phi_1$ into
$\phi =\phi_1 +\pi$ as can be done in the usual case. We will
call this type of singularity a "beak" (see Fig. 1b).

In the case of one real root of denominator $D$ the solution has the
same form as (\ref{r3}) but now e.g. the powers $\nu_2$ and $\nu_3$ are
complex conjugate.
The solution can be written as
\be
{r\over r_0} =(t^2 +1) \mid t - t_2 \mid ^{4{\rm Re}\, \nu_2}
\exp \left(4 \beta\, {\rm Im} \,\nu_2   \right)
\left( t-t_1 \right)^{2\nu_1}
\label{r1}
\ee
where $\beta = \tan^{-1} [ {\rm Im} t_2 /( t-{\rm Re} t_2 )]$.
The real root $\nu_1$ is negative, as is seen from eq. (\ref{nu1}) and
thus $r$ does not vanish in vicinity of such singular point. The
polarization direction field for this case is presented in Fig. 1c. This
type of singularity can be called a "comet".

One can estimate the density of different singular points in the
following way (see e.g. \cite{be,nn}). All singular points correspond to
the case when both $Q=0$ and $U=0$. The  number density of these points
is proportional to
\be
dQdU = {\rm det}
 \left( \begin{array}{cc}
Q_x & Q_y\\
U_x & U_y
\end{array}
\right) \, dx dy
\label{dqdu}
\ee
and thus the density is given by the average value of the determinate,
$d=Q_x U_y -Q_y U_x$. It can be easily checked that the saddle-type
singularity
takes place if $d>0$ that is the same condition as for normal
saddles in the field determined by the equation $dy/dx =Q/U$
see \cite{nn}.
It can be shown that saddles make 50\% of all singular points
$\langle n_s\rangle =0.5\langle n \rangle$, where $n$ is the number
 density of all singular points.
Calculations of the number of beaks and comets are more complicated and
can be found numerically.
According our estimates the surface densities for beaks and comets
are correspondingly $\langle n_b \rangle \approx 0.052\langle n \rangle$
and $\langle n_c\rangle \approx 0.448 \langle n \rangle$.
We note that the
probability of  appearance of saddles, beaks and comets for random choice
of $Q_x$, $Q_y$, $U_x$,$U_y$, is correspondingly $W_s=0.500$,
$W_b\approx 0.116$, $W_c\approx 0.384$.

\section{Types of perturbations field and types of
singular points.}
As we have already mentioned the
polarization of cosmic microwave radiation arises due to anisotropy of
the radiation field. It is a linear functional of the field and in
the case of scalar perturbations the only way to construct two dimensional
tensor quantity is to use second derivatives of a scalar function
$\Psi$, as is argued e.g. in ref. \cite{sz}, \cite{sz2}. The
matrix elements of traceless symmetric matrix (\ref{aqu}) are constructed
uniquely as
\be
a_{ij} = 2 \partial_i \partial_j \Psi - \delta_{ij}  \partial^2 \Psi
\label{mijs}
\ee
For the functions $Q$ and $U$ it gives:
\be
Q= (\partial^2_x -\partial^2_y) \Psi , \nonumber \\
U = 2\partial_x \partial_y \Psi
\label{qus}
\ee
In principle one may use also the invariant two-dimensional antisymmetric
(pseudo)-tensor $\epsilon_{ij}$ but parity conservation prevents from its
appearance in polarization matrix in the case of scalar perturbations. For
tensor perturbations there could be specific "external" directions in the
space and parity considerations do not prevent from using $\epsilon_{ij}$
in the matrix (\ref{aqu}) (see below). For this specific form (\ref{qus})
of polarization matrix there exist a particular (pseudo)scalar quantity
which vanishes in the absence of gravitational perturbations \cite{kks,zs}:
\be
B = \epsilon_{ij} \partial_k  \partial_i M_{kj}
\label{b}
\ee
Written explicitly it reads
\be
B = (\partial^2_x -\partial^2_y) U -  2\partial_x \partial_y Q
\label{b1}
\ee
It evidently vanishes for $Q$ and $U$ given by expressions (\ref{qus}).

In the case when tensor perturbations are present, there is more freedom
in polarization matrix $\bf a$  and the terms proportional to
$\epsilon_{ij}$ are permitted and the symmetric matrix $\bf a$
may be expressed through second derivatives of two independent
scalar functions:
\be
M_{ij} = 2 \partial_i \partial_j \Psi -\delta_{ij}  \partial^2 \Psi
     + \epsilon_{ik}\partial_k \partial_j  \Phi\,\,
\epsilon_{jk}\partial_k \partial_i \Phi
\label{mijt}
\ee
This is a general decomposion of symmetric traceless tensor in 2
dimensions (see e.g. \cite{kks}).

With inclusion of tensor perturbations the functions $Q$ and $U$ take
the form:
\be
Q= (\partial^2_x -\partial^2_y) \Psi +2\partial_x \partial_y \Phi ,
 \nonumber \\
U= 2\partial_x \partial_y \Psi - (\partial^2_x -\partial^2_y) \Phi
\label{qut}
\ee
The scalar $B$ is expressed through the fourth derivatives of $\Phi$
as following:
\be
B= \partial^4 \Phi
\label{bphi}
\ee

The difference between scalar and tensor perturbations appears only in
the fourth derivatives of  the generating scalar functions $\Phi$ and
$\Psi$, while the structure of their singularities is determined by the
third derivatives. Thus the types of the singularities are the same for
both types of perturbations.
There are statements in the literature that in the case of scalar
perturbations vector $\vec{n}$  cannot have curl, on the other hand tensor
perturbations do produce a curl, see \cite{ka}, section 4.

However the polarization tensor is
not a vector but a second rank tensor and direct analogy with a vector
field is not applicable. In the general case of singular points considered
above the curl is not equal to zero for any type of perturbations.
An explicit example of scalar generating function
$\Psi$ being a function only of $r=\sqrt{x^2+y^2}$ near the singularity
point shows
that the latter may be either center or knot. These points are absent in
our list of three presented above due to a specific degeneracy of
this example.

Thus to summarize, the singular points of the vector field $\vec n^{(1)}$,
which is the eigenvector of polarization matrix corresponding to the
direction of maximum polarization can be generically of the following
three types (see above): saddle and beak (with three separatrices), and
comet (one separatrix). In degenerate case, when some of the coefficients
or their combinations (like e.g. determinants) may be zero, then there
could be some other types singularities which we have not considered here.

\section{Statistics of singular points.}

The functions $Q$ and $U$ are usually assumed to be independent Gaussian
variables with equal dispersion \cite{be}:
\be
\langle Q^2 \rangle =\langle U^2 \rangle = \sigma_0^2
\label{qudisp}
\ee
Their first derivatives are also independent and non-correlated with the
functions with dispersion
\be
\langle Q_i Q_j \rangle =\langle U_iU_j \rangle =\delta_{ij} \sigma^2_1/2
\label{qiuidisp}
\ee
where $Q_i = \partial Q /\partial x^i$, etc.
All other correlators are zero.
However the fact that in the case of
scalar perturbations both functions $Q$ and $U$ as well as their
derivatives are determined by a single generating scalar function $\Psi$
imposes some conditions on the correlators of the second derivatives.
In particular the dispersions of the second derivatives $Q_{ij}$ and
$U_{ij}$ are not equal and these fields are correlated.
The list of nontrivial correlators is the following:
\be
\langle Q_{xx}^2 \rangle = \langle Q_{yy}^2 \rangle = \frac{7}{16}\sigma_2^2,
\,\,\,
\langle U_{xx}^2 \rangle = \langle U_{yy}^2 \rangle = \frac{5}{16}\sigma_2^2,
\nonumber \\
\langle Q_{xx} Q_{yy}\rangle=\langle Q_{xy}Q_{xy} \rangle
=\frac{1}{16}\sigma_2^2, \,\,\,
\langle U_{xx} U_{yy}\rangle=\langle U_{xy}U_{xy} \rangle
=\frac{3}{16}\sigma_2^2,
\\
\langle Q_{xx} U_{xy}\rangle=\langle Q_{xy}U_{xx} \rangle
=\frac{1}{16}\sigma_2^2, \,\,\
\langle Q_{yy} U_{xy}\rangle=\langle Q_{xy}U_{yy} \rangle
=-\frac{1}{16}\sigma_2^2, \,\,\
\nonumber \\
\langle Q_{xx} Q\rangle =\langle Q_{yy} Q\rangle=
\langle U_{xx} U\rangle = \langle U_{yy} U\rangle = - \sigma^2_1/2,
\nonumber
\label{corr}
\ee
where $\sigma_2^2\equiv \langle( Q_{xx}+Q_{yy})^2\rangle=
\langle(U_{xx}+U_{yy})^2\rangle$.
Note that the asymmetric correlators in the third line are non-vanishing.

These properties permit in principle to discriminate and detect a
contribution of tensor (or
vector) perturbations into polarization of CMB by measuring the dispersion
of second derivatives of the Stokes parameters. In particular for pure scalar
perturbations one should expect
\be
{\langle Q_{xx}^2 \rangle \over \langle U_{xx}^2 \rangle } =
{\langle Q_{yy}^2 \rangle \over \langle U_{yy}^2 \rangle } = {7/5}, \,\,\
\langle (U_{xx}-U_{yy})^2\rangle = 4 \langle Q_{xy}^2 \rangle
\label{qoveru}
\ee
A deviation from this number would indicate a contribution from other,
different from scalar, types of perturbations. The last equality in (27)
corresponds to $B=0$, see Eq.(20). The property $B=0$ as a test for the
absence of scalar perturbations was indicated in previous works, see for
 example \cite{zs}.

An interesting quantity which allows one to relate global characteristics
of random field to local properties is the Euler characteristic, $\chi_E$,
see \cite{cdp}. As it was noted in ref. \cite{mw}
this value is closely linked to the critical value
of the amplitude of polarization, $P=\sqrt{Q^2 +U^2}$, for which the
regions of high polarization percolate.
According to this [14] the percolation begins at the amplitude of
polarization corresponding to $\chi_E=0$.

This critical amplitude was estimated in ref. \cite{nn} where it was
found that percolation takes place for $p=1$, where $p = P/\sigma_0$
is the dimensionless amplitude of polarization with unit dispersion.
We estimated this quantity in somewhat different way than it was done in
ref. \cite{nn}. In the 2D case the required value is defined by an
equation
\be
\chi_E\propto \langle p_{xx}+p_{yy}\rangle p\exp(-p^2/2)\propto (p^2-1)
\exp(-p^2/2)
\ee
that is identical to result obtained by ref. \cite{nn}.
Though the statistical distribution we used is different
from the one of ref. \cite{nn}, we got the same result: percolation
occurs at $p=1$. Let us remind that for the 2D Gaussian fields
$\chi_E\propto p\exp(-p^2/2)$ and percolation occurs at $p=0$.
As it is following from (17) - (22) the same results are valid for
all three types of perturbations of polarization field.
This topics will be discussed in more detail in \cite{ddnn}.

\acknowledgments

This work was supported in part by Danmarks Grundforskningsfond through
its funding of the Theoretical Astrophysical Center (TAC), the Danish
Natural Science Research Council through grant No 9401635, and in part
by NSF-NATO fellowship (DGE-9710914) and by NSF EPSCOR program.

\begin{figure}
\epsfbox{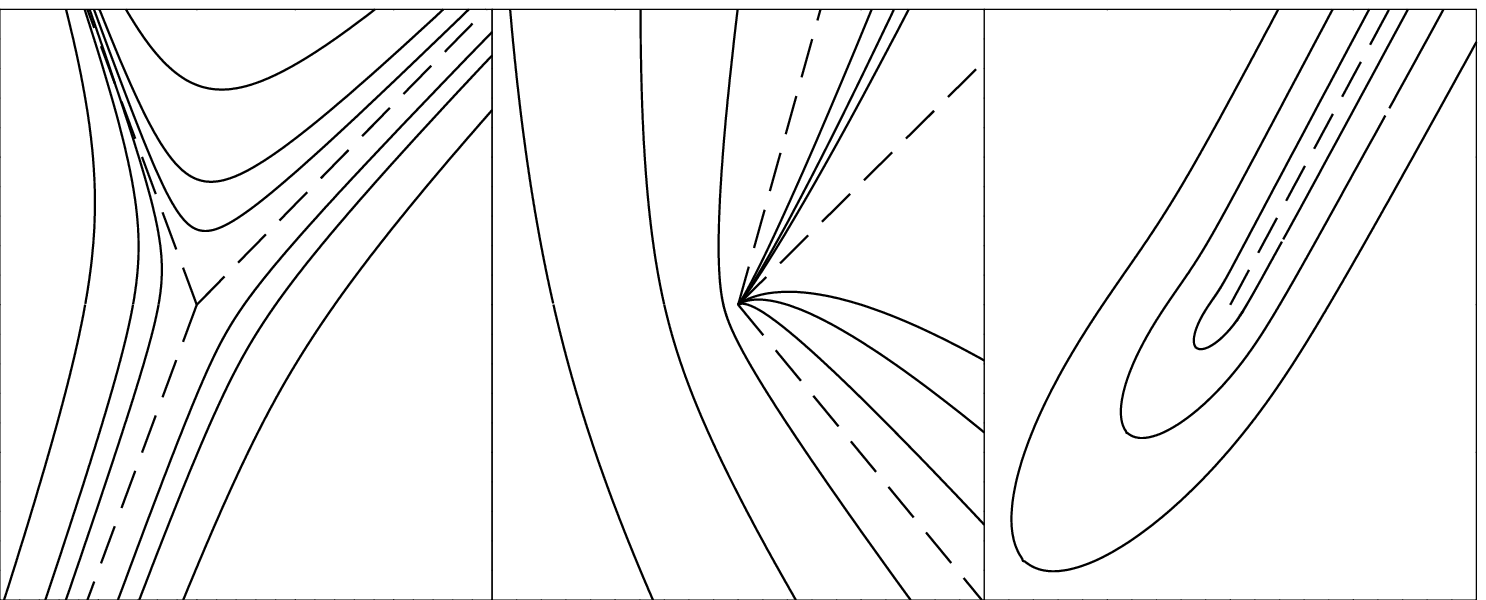}
\vspace{1.cm}
\caption{Integral curves for three different 
types of singular points: 
a. saddle, b. beak, \& c. comet. Long dashed lines show 
peculiar solutions (separatrises).} 
\label{fig1}
\end{figure}

\end{document}